# Using detrended deconvolution foreign exchange network to identify currency status

Pengfei Xi[1, *], Shiyang Lai[1], Xueying Wang[1], Weiqiang Huang[1]

1. School of Business Administration, Northeastern University, P.R. China;

**Abstract**: This article proposed a hybrid detrended deconvolution foreign exchange network construction method (DDFEN), which combined the detrended cross-correlation analysis coefficient (DCCC) and the network deconvolution method together. DDFEN is designed to reveal the 'true' correlation of currencies by filtering indirect effects in the foreign exchange networks (FXNs). The empirical results show that DDFEN can reflect the change of currency status in the long term and also perform more stable than traditional network construction methods.

## 1. Introduction

In recent years, using FXNs to analyze the international currency market has become a thriving topic in economics research. However, due to the complex nature of the foreign exchange (FX) relationships, it is almost impossible to operate practical analysis on the original FX correlation networks. To overcome this, Mcdonald et al. (2005) innovatively proposed a minimum spanning tree (MST) based FXNs construction method helping to simplify the overcomplicate FX linkage system, which has caught a lot of researcher's attention (Kwapień et al., 2009; Li and Liao, 2019; Sinha and Kovur, 2014). But in fact, the MST based method just give an oversimplified solution to this problem, it only keeps the connection of each currency to its most relevant one and ignore all of the others, which inevitably cause much information loss. To further improve the construction method of FXNs, this paper introduced DDFEN to derive more diverse and realistic currency relationships.

To achieve our purpose, the two main issues we need to concern is finding the right correlation level of each two currency pairs and omitting those links caused by indirect effects. Inspired by the work of Li et al. (2018), who argues that DCCC is a better way to calculate the correlation between currency pairs, we employed DCCC to get the original correlation network in order to reflect more reality FX correlation level. But the correlation shown between two currencies may be due to other currencies associated with both of them (Figure 1), in this regard, this kind of relationship is indirect and not 'truly exist'. To omit these links in FXNs, the technical approach we adopt here to filter out this connection is network deconvolution. Feizi et al. (2013) introduced an efficient and scalable algorithm for deconvolving an observed network, which provided us with a prototype for weakening the indirect effects in FXN. To demonstrate our method is applicable, we utilized the past ten years exchange rate data to construct MST-based FXNs and DDFEN-based FXNs, using them to identify the change of CNY's status. Our empirical results show that, compared with traditional construction methods based FXNs, most commonly used network analysis techniques can present more stable analysis results in DDFEN-based FXNs.

---

* Corresponding author
  *Email address:* xpf2017@stumail.neu.edu.cn (Pengfei Xi)

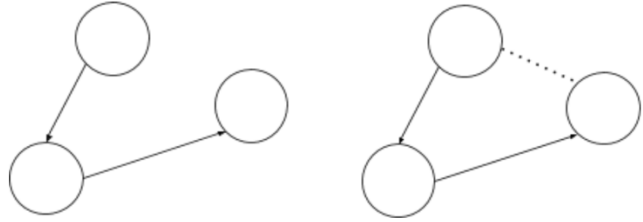

**Figure 1.** Example of indirect relationship in FXNs

The remainder of this paper is organized as follows. Section 2 describes the details of the algorithm. Section 3 reports the empirical results, and the last section presents the conclusions.

## 2. Methodology and data description

Our methodology can be separate into three main parts, i.e., calculating the FX correlation matrix, adjusting the weight by network deconvolution algorithm, and finding the maximum threshold that guarantees full network connectivity.

With DCCC, we calculate the correlation between two currency pairs $\{i_n\}$ and $\{j_n\}$ for a window size $w$ as

$$\text{Corr}_{DCCC}(i,j,w) = \frac{F^2_{DCCA}}{F_{DFA,i}(w)F_{DFA,j}(w)}$$

where $n$ represents the length of two currency pairs series. $F_{DCCA}(w)$, $F_{DFA,i}(w)$ and $F_{DFA,j}(w)$ are the detrended covariance between profiles $\{I_n\}$ and $\{J_n\}$ detrended variances of $\{I_n\}$ and detrended variances of $\{J_n\}$, respectively.

The profile of time series $\{i_n\}$ and $j_n\}$ is defined as

$$I_n = \sum_{\alpha=1}^{n}(i_\alpha - \hat{i})$$

$$J_n = \sum_{\alpha=1}^{n}(j_\alpha - \hat{j})$$

where $\hat{i}$ and $\hat{j}$ represent the mean value of each time series.

We divide the profile of $\{i_n\}$ and $j_n\}$ into the sliding window of size $w$. In every position $\beta$ where the window slides, a time trend has been simply fitted by a linear regression, thus, fitted values (by least square method) of $I_{\gamma,\beta}$ and $J_{\gamma,\beta}$ for $\beta \leq \gamma \leq \beta + w - 1$ can be obtained easily. Based on all of these information, detrended variance is then defined as

$$f^2_{DFA,i}(w,\beta) = \frac{\sum_{\gamma=\beta}^{\beta+w-1}(I_\gamma - \hat{I}_{\gamma,\beta})^2}{w-1}$$

$$f^2_{DFA,j}(w,\beta) = \frac{\sum_{\gamma=\beta}^{\beta+w-1}(J_\gamma - \hat{J}_{\gamma,\beta})^2}{w-1}$$

After that, the detrended variance of $\{I_n\}$ and $\{J_n\}$, namely, $F_{DFA,i}(w)$ and $F_{DFA,j}(w)$, can be achieved by calculating the average of $f^2_{DFA,i}(w,\beta)$ and $f^2_{DFA,j}(w,\beta)$.

$$F^2_{DFA,i}(w) = \frac{\sum_{\beta=1}^{n-w+1} f^2_{DFA,i}(w,\beta)}{n-w}$$

$$F^2_{DFA,j}(w) = \frac{\sum_{\beta=1}^{n-w+1} f^2_{DFA,j}(w,\beta)}{n-w}$$

Similarly, the detrended covariance of profiles is defined as

$$f^2_{DCCA}(w,\beta) = \frac{\sum_{\gamma=\beta}^{\beta+w-1}(I_\gamma - I_{\gamma,\beta})(J_\gamma - J_{\gamma,\beta})}{w-1}$$

and the detrended covariance between profiles $\{I_n\}$ and $\{J_n\}$ is calculated by

$$F^2_{DCCA}(w) = \frac{\sum_{\beta=1}^{n-w+1} f^2_{DCCA}(w,\beta)}{n-w}.$$

By using DCCC for the input currency pairs matrix, we looped over all the currency pairs combinations and finally achieved the correlation matrix $M(w)$, for a window size $w$

$$M_{i,j}(w) = Corr_{DCCC}(i,j,w)$$

Having obtained the currency correlation matrix $M(w)$ through DCCC, we use network deconvolution algorithm to construct direct currency relationship network. The specific deconvolution steps as follows:

Decompose the currency correlation matrix $M(w)$ to matrix of eigenvectors $U$ and diagonal

matrix of eigenvalues (composed by $i^{th}$ eigenvalue $\lambda_i$) such that

$$M(w)=U\begin{pmatrix} \lambda_1 & 0 & \cdots & 0 \\ 0 & \lambda_2 & & \\ \vdots & & \ddots & \\ 0 & & & \lambda_n \end{pmatrix}U^{-1}.$$

Form a new diagonal matrix whose $i^{th}$ diagonal component is

$$\lambda_d = \frac{\lambda}{1+\lambda}$$

Then the deconvolution correlation matrix $M_d$ is

$$M_d=U\begin{pmatrix} \lambda_1^d & 0 & \cdots & 0 \\ 0 & \lambda_2^d & & \\ \vdots & & \ddots & \\ 0 & & & \lambda_n^d \end{pmatrix}U^{-1} = U\begin{pmatrix} \frac{\lambda_1}{1+\lambda_1} & 0 & \cdots & 0 \\ 0 & \frac{\lambda_2}{1+\lambda_2} & & \\ \vdots & & \ddots & \\ 0 & & & \frac{\lambda_n}{1+\lambda_n} \end{pmatrix}U^{-1}$$

To omit those links that weaken by deconvolution and make sure that all nodes in the network are not isolated, the way we determine the threshold can be summarized as follows:

First, find the maximum values in each row of $M_d$, $\sigma = \{\sigma_1, \sigma_2, ..., \sigma_n\}$, where

$$\sigma_i = \max_j M_d(i,j)$$

Then, get the minimum value in $\sigma$ and denote it as $\sigma_{\min}$.

After that, calculate the percentage filtering threshold corresponding to $\sigma_{\min}$:

$$\theta = \frac{l(\{M_{d(i,j)} \mid M_{d(i,j)} \geq \sigma_{\min}\})}{l(\{M_{d(i,j)}\})}$$

where $l(\{M_{d(i,j)} \mid M_{d(i,j)} \geq \sigma_{\min}\})$ is the size of the subset $\{M_{d(i,j)} \mid M_{d(i,j)} \geq \sigma_{\min}\}$, $l(\{M_{d(i,j)}\})$ is the size of the subset $\{M_{d(i,j)}\}$. Finally, using the percentage threshold, we keep the top $\theta$ of the correlation values in $M_d$ and obtain the DDFEN-based FXN, $M_d$.

According to some recent empirical studies (Dave and Kobayashi, 2018; Du and Zhang, 2018; Chan, 2017), the status of CNY is rising obviously in the last ten years. To test the ability of DDFEN on detecting the dynamic change of CNY's status, 70 countries' FX rate (quoted against USD) daily data of historical price-postings from March 27, 2010, to January 30, 2020, was chosen in this article as the dataset (from wind co database).

## 3. Empirical results

To compare with the traditional method, the empirical results of DDFEN-based FXNs and MST based FXNs are both presented in this section. We selected three important time markers which have been considered as remarkable events implying the rising of CNY's status (Dave and Kobayashi, 2018; Zhao, 2016) to test the performance of two methods, i.e., the implementation of the official document of China's "One Belt and One Road" policy (abbreviated as "The B&R") on March 28, 2015, the establishment of the New Development Bank (abbreviated as "NDB") on March 27, 2013, and the holding of the Shanghai Cooperation Organization (abbreviated as "SCO") on June 15, 2011. In this experiment, a window length of $w$=200 days was rolled forward at t =60 days intervals. To measure the change of the status of CNY, four popular network indexes, which can reflect the importance of currencies in FXNs (Jo et al., 2018; Li et al., 2018), were chosen for comparison (Table 1).

**Table 1.** Four network indexes for node status

| Index | Formula | Description |
|---|---|---|
| Weighted Degree Centrality | $C_w(i) = \sum_{j=1}^{g} x_{i,j} (i \neq j)$ | Sum of the weighted edges connected to node $i$ except itself. |
| Authority | $A(i) = a_i = \sum_{i=1}^{n} x_{ij} h_k$ <br> $h_k = \sum_{i=1}^{n} x_{i,j} a_k$ | The sum of the hub scores of node $i$'s neighborhoods. |
| Closeness Centrality | $C_C(i) = \dfrac{n-1}{\sum_{j \neq i} d_{i,j}}$ | The reciprocal of the average of the distances from node $i$ to all other nodes. |
| Betweenness Centrality | $C_B(k) = \sum_{i \neq k \neq j} \dfrac{g_{i,j}(k)}{g_{i,j}}$ | The $g_{i,j}(k)$ is the number of shortest paths between $n_i$ and $n_j$ that pass through $n_k$. $g_{i,j}$ respects the number of shortest paths between $n_i$ and $n_j$. |

Note: $x_{i,j} = M_{d(i,j)}$, $V$ is the collection of all nodes in $M_d$.

Figure 1 presents the result of the Weighted Degree Centrality. The step line in the above subgraph (DDFEN) has a clear and stable growth trend. Meanwhile, its fluctuations coincide with the time markers we have chosen. On the contrary, the result from MST (subgraph below) doesn't show us any trends, and the changes of the step line are too random to say its fluctuation corresponding to our time markers. The results of other indexes (attached in the appendix A) are similar, and they show that DDFEN-based FXNs can reflect the status changes of CNY in most commonly used network analysis techniques. Comparing with MST, DDFEN has a clear advantage in accurate recognition of currency status.

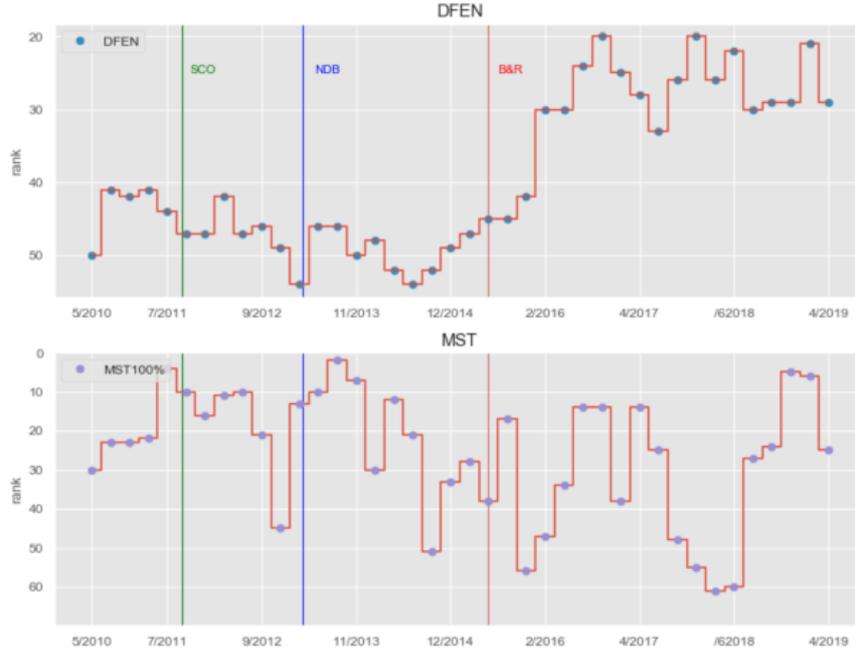

**Figure 2.** CNY ranked by Weighted Degree Centrality

To further compare the stability of DDFEN and MST, we analyze the fluctuation level in different indexes. We apply the least square method to get the detrended volatility, $\sqrt{\dfrac{\sum_{i=1}^{n}(rank_i - \hat{rank}_i)}{n}}$ ($rank_i$ is the value get from index, $\hat{rank}_i$ is the fitted value from general linear regression, $n$ is the number of time points). According to the results of our stability test (Table 2), the volatility of DDFEN is significantly lower than MST in each index, which indicates the stability of DDFEN in identifying currency status is much better than the traditional method.

**Table 2.** Fluctuation in CNY rankings over time

|  | Weighted Degree | Authority | Closeness Centrality | Betweenness Centrality |
| --- | --- | --- | --- | --- |
| MST | 15.583 | 20.182 | 17.749 | 18.198 |
| DDFEN | 7.030 | 6.158 | 14.267 | 14.487 |

## 4. Conclusion

In this article, we introduced a hybrid FXNs construction method, namely, DDFEN, which combined the DCCC correlation matrix and the network deconvolution method together, to identify the dynamic changes of currency's status. By comparison, we argue that MST, the mainstream FXN construction technique, tends to oversimplify the correlations between each currency, which will lead to the loss of valuable information. In contrast, much closer to the real currency market, DDFEN regards the FX rate as non-stationary time series and using DCCC to describe the correlation between FX rates. Besides, DDFEN only filters out indirect relationships between currencies and allow more connected paths exist between two currencies. Our empirical results show DDFEN-based FXNs are more stable than MST-based FXNs, and the dynamic change of CNY's status identified by DDFEN is quite close to general recognition. For scholars who want to adopt network analysis to macroeconomics, social sciences, and information science, DDFEN is equally applicable. It can be regarded as an effective approach to construct non-stationary sequence correlation networks.

# Appendix A

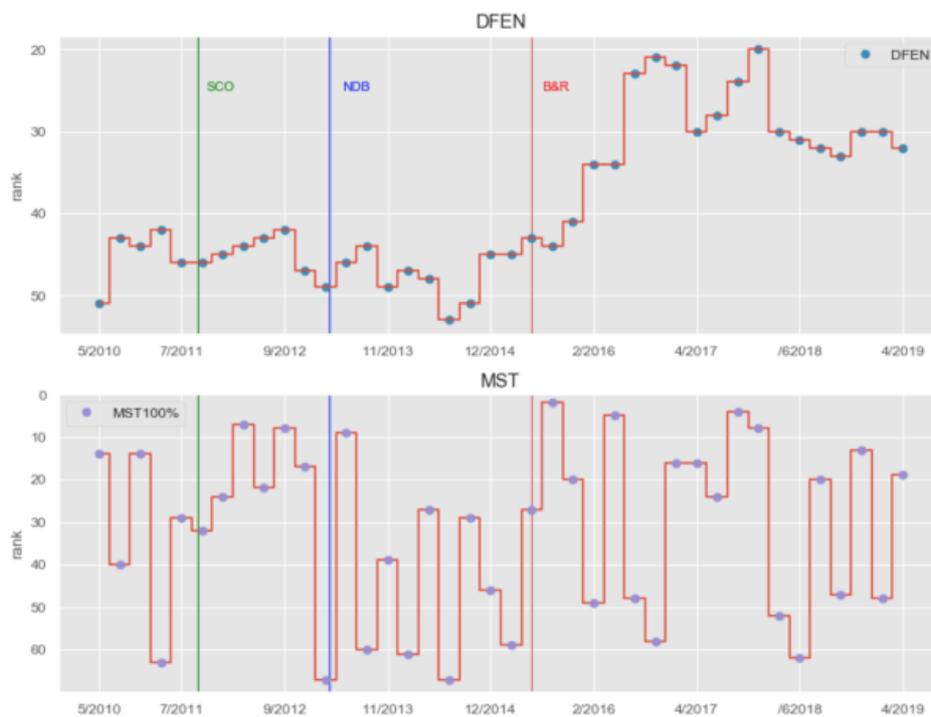

**Table A1.** CNY ranked by Authority

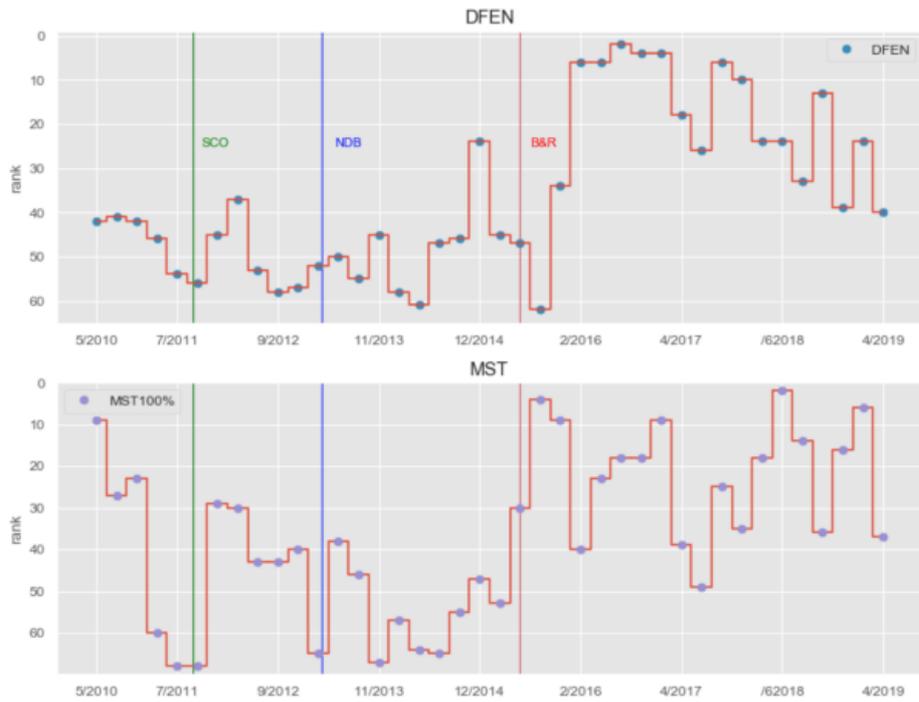

**Table A2.** CNY ranked by Closeness Centrality

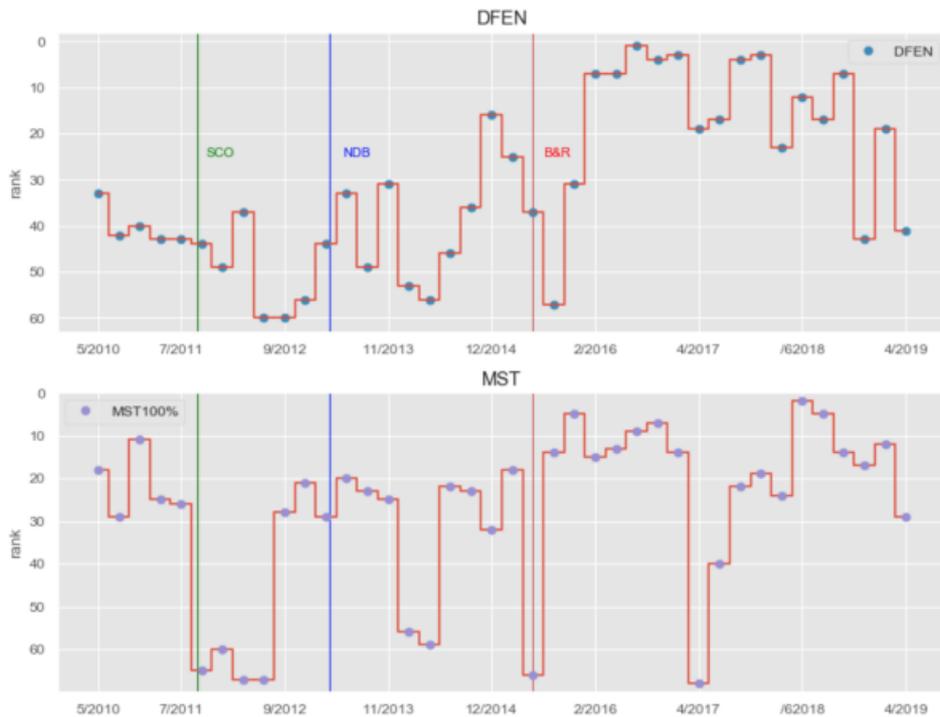

**Table A3.** CNY ranked by Betweenness Centrality